%% file: mnras_template.tex
\documentclass[fleqn,usenatbib]{mnras}

\usepackage[T1]{fontenc}
\usepackage{amsfonts}
\usepackage{algorithm}
\usepackage{algpseudocode}

\DeclareRobustCommand{\VAN}[3]{#2}
\let\VANthebibliography\thebibliography
\def\thebibliography{\DeclareRobustCommand{\VAN}[3]{##3}\VANthebibliography}

\usepackage{graphicx}	
\usepackage{amsmath}	
\usepackage{cleveref}

\title[Accelerated NS with $\beta$-flows]{Accelerated nested sampling with posterior repartitioning and $\beta$-flows for gravitational waves}

\author[]{
Metha Prathaban,$^{1,2,3}$\thanks{E-mail: myp23@cam.ac.uk}
Harry Bevins,$^{1,2}$
Will Handley,$^{1,2,4}$
\\
$^{1}$Kavli Institute for Cosmology, Madingley Road, Cambridge CB3 0HA, UK\\
$^{2}$Astrophysics Group, Cavendish Laboratory, J.J. Thomson Avenue, Cambridge CB3 0HE, UK\\
$^{3}$Pembroke College, Trumpington Street, Cambridge CB2 1RF, UK \\
$^{4}$Gonville \& Caius College, Trinity Street, Cambridge CB2 1TA, UK
}

\date{Accepted XXX. Received YYY; in original form ZZZ}

\pubyear{\the\year{}}

\begin{document}
\label{firstpage}
\pagerange{\pageref{firstpage}--\pageref{lastpage}}
\maketitle

\begin{abstract}
There is an ever-growing need in the gravitational wave community for fast and reliable inference methods, accompanied by an informative error bar. Nested sampling satisfies the last two requirements, but its computational cost can become prohibitive when using the most accurate waveform models. In this paper, we demonstrate the acceleration of nested sampling using a technique called posterior repartitioning. This method leverages nested sampling's unique ability to separate prior and likelihood contributions at the algorithmic level. Specifically, we define a `repartitioned prior' informed by the posterior from a low-resolution run. To construct this repartitioned prior, we use a $\beta$-flow, a novel type of conditional normalizing flow designed to better learn deep tail probabilities. $\beta$-flows are trained on the entire nested sampling run and conditioned on an inverse temperature $\beta$. Applying our methods to simulated and real binary black hole mergers, we demonstrate how they can reduce the number of likelihood evaluations required for a given evidence precision by up to an order of magnitude, enabling faster model comparison and parameter estimation. Furthermore, we highlight the robustness of using $\beta$-flows over standard normalizing flows for posterior repartitioning. Notably, $\beta$-flows are able to recover posteriors and evidences which are generally consistent with those from traditional nested sampling, even in cases where standard normalizing flows fail.


\end{abstract}




\section{Introduction}

Nested sampling (NS)~\citep{Skilling2006NS} is a Bayesian inference tool widely used across the physical sciences, including in the analysis of gravitational wave (GW) data~\citep{Ashton2022NSReview, Thrane_2019, Veitch_LAL, bilby_paper}. Unlike many Bayesian inference algorithms that focus solely on approximating the posterior distribution from a given likelihood and prior, nested sampling first evaluates the Bayesian evidence. This evidence, obtained by evaluating an integral over the parameter space, is essential for model comparison and tension quantification. Samples from the normalized posterior can then be drawn as a byproduct of this calculation.

While the ability to compute evidences is a key advantage, nested sampling can be slower than alternative posterior samplers, such as Metropolis-Hastings~\citep{metropolis1953, hastings1970}. This challenge is particularly pronounced in the analyses of compact binary coalescences (CBCs) in gravitational wave data, where the use of high-fidelity waveform models or models incorporating additional physics can make likelihood evaluations prohibitively expensive. Even for faster waveform models, standard nested sampling for third-generation (3G) GW detectors is expected to be impractically slow~\citep{VeitchAccelerationReview}. Consequently, reducing the wall-time for inference has been the focus of significant research efforts~\citep{DINGO, ROQ, ROQ2, ROQ3, Multibanding1, Multibanding2, TL_relativebinning, relativebinning2, relativebinning3, relativebinning4, likelihood_reweighting, TemperedImportanceSampling}.



Several methods have been proposed to accelerate the core NS algorithm~\citep{Supernest, DynamicNS}, with one promising solution being posterior repartitioning (PR)~\citep{PR1}. Originally introduced to solve the problem of unrepresentative priors, this approach takes advantages of NS's unique ability in distinguishing between the prior and the likelihood, by sampling from the prior, $\pi$, subject to the hard likelihood constraint, $\mathcal{L}$. Other techniques, such as Hamiltonian Monte Carlo~\citep{HMC1, HMC2} and Metropolis-Hastings, are only sensitive to the product of the two. PR works by redistributing parts of the likelihood into the prior that NS sees, thereby reducing the number of iterations of the algorithm required for convergence~\citep{Supernest}. The main difficulty lies in defining the optimal prior for this purpose.


Normalizing flows (NFs) offer a promising approach to addressing this. These versatile generative modelling tools have been widely adopted in the scientific community for tasks ranging from performing efficient joint analyses~\citep{Bevins2022margarine1, Bevins2023margarine2} to evaluating Bayesian statistics like the Kullback-Leibler divergence in a marginal framework~\citep{Bevins2023margarine2, Pochinda2023Constraints, Gessey-Jones2024Constraints}, as region samplers in the nested sampling algorithm~\citep{Williams2021Nessai}, as proposals for importance sampling and MCMC methods~\citep{Papamakarios2015ImportanceSampling, Paige2016SMC, Matthews2022SMC} and as a foundation for Simulation Based Inference~\citep{Fan2012NLE, Papamakarios2016NPE}, among others. 

Importantly, they can also be used to define non-trivial priors~\citep{Alsing2022anyprior, Bevins2023margarine2}, making them ideal candidates for use as repartitioned priors in PR to speed up NS. Central to the success of this application of normalizing flows, and indeed of all the above applications, is the accuracy of the flow in representing the distribution it aims to learn. In this paper, we will demonstrate empirically that the accuracy of commonly used normalizing flow architectures is often poor in the tails of the distribution. We introduce $\beta-$flows, which are trained on the whole nested sampling run and conditioned on an inverse temperature $\beta$, analogous to the inverse temperature in statistical mechanics. Since NS has deep tails, $\beta$-flows are able to better learn the tails of target distributions. We show that replacing standard normalizing flows with $\beta$-flows can lead to improvements in the runtime and robustness of PR-accelerated NS.

In the following section, we lay out the necessary background. We then introduce $\beta$-flows and describe the methodology used in our analyses in Section~\ref{methods}, and present and discuss our results in Section~\ref{results}. Finally, conclusions are presented in Section~\ref{conclusions}. 

\section{Background}

Section~\ref{sec:NS_intro} provides a brief overview of the key concepts of nested sampling and establishes notation. For a more detailed review, readers are directed to~\cite{Skilling2006NS} and~\cite{Ashton2022NSReview} for general information on NS, and to~\cite{Polychord1} for specifics about \textsc{PolyChord}, the NS implementation used in this work. Sections~\ref{sec:runtime} and~\ref{sec:PR} provide background on the runtime of NS and outline posterior repartitioning, introducing key aspects that extend beyond the standard nested sampling framework. 

\subsection{Nested sampling and Bayesian inference}\label{sec:NS_intro}

\begin{figure}\vspace{-50pt}
\centering
\def\svgwidth{1.2\columnwidth}
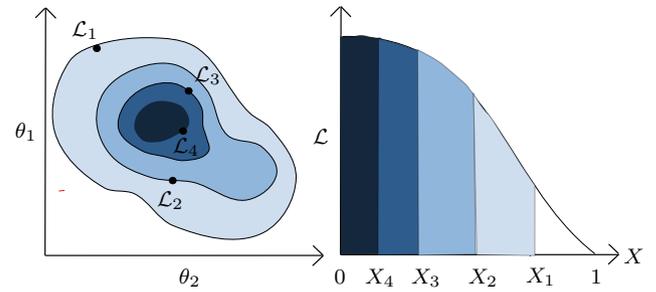
\vspace{-260pt}
\caption{Schematic of a nested sampling run. Each dead point defines an iso-likelihood contour in the parameter space (left), which then encloses a certain fractional prior volume (right). As the points compress towards the peak of the likelihood, they enclose smaller and smaller fractional volumes.}
\label{fig1}
\end{figure}

The nested sampling algorithm, first proposed by~\cite{Skilling2006NS}, is a technique whose primary goal is to calculate the evidence term in Bayes' theorem. Given some model $\mathcal{M}$ and observed data $D$, Bayes' theorem enables us to relate the posterior probability of a set of parameters $\theta$ to the likelihood, $\mathcal{L}$, of $D$ given $\theta$ and the prior probability, $\pi$, of $\theta$ given $\mathcal{M}$

\begin{equation}
    P(\theta | D, \mathcal{M}) = \frac{P(D | \theta, \mathcal{M}) P(\theta | \mathcal{M})}{P(D | \mathcal{M})} = \frac{\mathcal{L}(D | \theta)\pi(\theta)}{\mathcal{Z}}.
\end{equation}
In general, the evidence, $\mathcal{Z}$, is a many dimensional integral over the parameter space:



\begin{equation}
    \mathcal{Z} = \int \mathcal{L}(\theta) \pi(\theta) d\theta.
\end{equation}
The innovation of NS is in transforming this into a one dimensional problem, by defining the integral in terms of the fractional prior volume enclosed by a given iso-likelihood contour at $\mathcal{L}^\ast$ in the parameter space:

\begin{equation}
    X(\mathcal{L}^\ast) = \int_{\mathcal{L}>\mathcal{L}^\ast} \pi(\theta)d\theta.
\end{equation}
In this way, the integral may be written as:

\begin{equation}\label{evidenceX}
    \mathcal{Z} = \int \mathcal{L}(X) dX.
\end{equation}
The NS algorithm begins by populating the prior with a set `live points'. At each iteration $i$, the live point with the lowest likelihood is deleted, and a new live point is sampled from the prior with the constraint that its likelihood, $\mathcal{L}$, must be higher than that of the deleted point, $\mathcal{L}^\ast$. The algorithm terminates once some set stopping criterion is satisfied, at which point the evidence may be estimated as a weighted sum over the deleted, or `dead', points; the weights correspond to the fractional prior volumes of the `shells' enclosed between successive dead points, $w_i = \Delta X_i = X_{i-1} - X_i$. A schematic of this is shown in Figure~\ref{fig1}.

\begin{equation}\label{evidence_weightedsum}
    \mathcal{Z} = \sum_{\textrm{dead points}} \mathcal{L}_i w_i.
\end{equation}
The posterior weights of the dead points are given by
\begin{equation}\label{dead_posteriorweight}
    p_i = \frac{w_i \mathcal{L}_i}{\mathcal{Z}}.
\end{equation}

\subsection{Runtime and acceleration of NS }\label{sec:runtime}

The nested sampling algorithm typically terminates when the estimated evidence remaining in the live points is below some set fraction of the accumulated evidence so far. The total convergence time may be expressed as~\citep{Supernest}:

\begin{equation}\label{runtime}
    T \propto  T_\mathcal{L} \times f_\textrm{sampler} \times D_{\textrm{KL}} \times n_\textrm{live},
\end{equation}
where $n_\textrm{live}$ is the number of live points, $T_\mathcal{L}$ is the time taken for a single likelihood evaluation, $f_\textrm{sampler}$ encapsulates the average number of calls to the likelihood function to choose a new live point, dependent on the sampler implementation, and $D_\mathrm{KL}$ is the Kullback-Liebler divergence, representing the amount of compression from prior to posterior. This is defined as:

\begin{equation}
    \mathcal{D}_{\textrm{KL}} = \int \mathcal{P}(\theta) \textrm{ln} \frac{\mathcal{P}(\theta)}{\pi(\theta)} d\theta.
\end{equation}
Historically in gravitational wave analyses, much of the efforts in bringing down the wall-time for inference has focused on the $T_\mathcal{L}$ term, which involves developing faster waveform models through various approximations~\citep{TL_phenomD, IMRPhenomXPHM, TL_fastmodel, TL_ROQreview, TL_multibandinterpolation, TL_relativebinning}. Meanwhile, the nested sampling community has emphasized developing samplers which reduce the $f_{\textrm{sampler}}$ term~\citep{Polychord1, Polychord2, NS_multinest, NS_multinest2, NS_cosmonest, NS_cosmonest2, NS_dynesty, NS_dypolychord, NS_ultranest, Williams2021Nessai, NS1, NS2, NS3, NS4, NS5, NS6, NS7, NS8, NS9, NS10, NS11, NS12}. The aim of this paper is to accelerate NS by taking advantage of the runtime's dependence on the KL divergence term. 

The KL divergence is particularly important because it appears again in the uncertainty of the accumulated evidence. We may express the uncertainty in $\textrm{log}\mathcal{Z}$ as

\begin{equation}\label{uncertainty}
    \sigma_{\mathrm{log}\mathcal{Z}} \propto \sqrt{D_\textrm{KL} / n_\textrm{live}}.
\end{equation}
For a fixed uncertainty $\sigma$, $n_\mathrm{live}$ is directly proportional to $D_\mathrm{KL}$: a lower KL divergence allows for fewer live points, further reducing the time to convergence without sacrificing precision. In this sense, the precision-normalized runtime of NS has a quadratic dependence on the KL divergence. Thus, an effective way to accelerate NS is to reduce the amount of compression from prior to posterior. 

In practice, one way to achieve this is to first perform a low resolution pass of NS to identify roughly the region of the parameter space where the posterior lies. Then, a narrower box prior can be set in this region for high resolution pass. The tighter prior used in the second pass reduces the KL divergence between the prior and posterior. However, since the prior has changed, the evidence from the second pass will not be the desired evidence. For simple box priors, this can be corrected after the run by multiplying the second pass's evidence by the ratio of the prior volumes to recover the original evidence. For more details and an application of this method, see, for example,~\cite{anstey}. 

This method can be further improved by training a normalizing flow (NF) on the rough posterior from the low resolution pass and using this as the new prior for the high resolution pass, instead of a simple box. NFs are generative models which transform a base distribution onto a more complex one by learning a series of invertible mappings between the two. For further details on normalizing flows, readers are referred to~\cite{NF_review} for an introduction and review of the current methods, and to~\cite{Bevins2022margarine1, Bevins2023margarine2} for details on \textsc{margarine}, the \textsc{python} package used to train the normalizing flows in this work. 

However, when using the output of trained flows as the new proposal, it is no longer trivial to correct the evidence exactly. Other techniques must be employed to address this issue.

\subsection{Posterior repartitioning}\label{sec:PR}

Many sampling algorithms, such as Metropolis Hastings~\citep{metropolis1953, hastings1970} and Hamiltonian Monte Carlo~\citep{HMC1, HMC2}, are sensitive only to the product of the likelihood and prior\footnote{This is known as the `unnormalized posterior' and is in fact the joint distribution. It is this joint distribution that is used, for example, in the Metropolis acceptance ratio.}. Nested sampling on the other hand, in ``sampling from the prior, $\pi$, subject to the hard likelihood constraint, $\mathcal{L}$'', uniquely distinguishes between the two~\citep{Supernest}. Given that the evidence and posterior only depend on $\mathcal{L} \times \pi$, it follows that we are free to repartition the prior and likelihood that nested sampling sees in any way, as long as their product remains the same:

\begin{align}
    & \tilde{\mathcal{L}}(\theta) \tilde{\pi}(\theta) = \mathcal{L}(\theta) \pi(\theta)  \\
     &\implies \tilde{\mathcal{Z}} = \int \tilde{\mathcal{L}}(\theta)\tilde{\pi}(\theta) d\theta = \int \mathcal{L}(\theta) \pi (\theta) d\theta = \mathcal{Z}; \\
     &\implies\tilde{\mathcal{P}}(\theta) = \frac{\tilde{\mathcal{L}}(\theta)\tilde{\pi}(\theta)}{\tilde{\mathcal{Z}}} = \frac{\mathcal{L}(\theta) \pi (\theta)}{ \mathcal{Z}} = \mathcal{P}(\theta).
\end{align}

This concept of `posterior repartitioning' (PR) was originally introduced by~\cite{PR1, PR2} as a way to tackle problems where the prior may be unrepresentative. They pioneered a specific implementation of this called `power posterior repartitioning' (PPR), where the original prior is raised to a power $\beta$, where $\beta$ is treated as a hyperparameter which is sampled over during the run. This new adaptive prior can then widen itself at runtime if the original prior was indeed unrepresentative. Although conceived for the purposes of robustness, the same fundamental ideas can be applied to speed up NS. As explained in Section~\ref{sec:runtime}, the inference time depends on the amount of compression between prior and posterior. Hence, moving portions of the likelihood into the nested sampling prior such that it is closer to the posterior means a smaller KL divergence and a faster run. Crucially, the product of the likelihood and prior remaining the same means we can get the correct evidences out in the first instance, bypassing the need to correct them by a prior volume factor as in~\cite{anstey}. These techniques have been applied in~\cite{Supernest} to accelerate NS, although not with $\beta$-flows. 

\section{Methods}\label{methods}

Putting the above pieces together, we can accelerate NS by running a low resolution pass first, training a NF on this and then using the NF as the prior for a second, higher resolution run. We also alter the likelihood for this second run, in accordance with PR, so that
\begin{align}\label{PR_prior}
\pi^\ast & = \mathrm{NF}(\theta) \\ 
\mathcal{L}^\ast & = \frac{\mathcal{L}(\theta)\pi(\theta)}{\mathrm{NF}(\theta)}\label{PR_likelihood},
\end{align}
where $\mathrm{NF}(\theta)$ is the probability of $\theta$ predicted by the NF and $\mathcal{L}$ and $\pi$ represent the original likelihood and prior respectively. 

\begin{figure}
    \centering
    \includegraphics{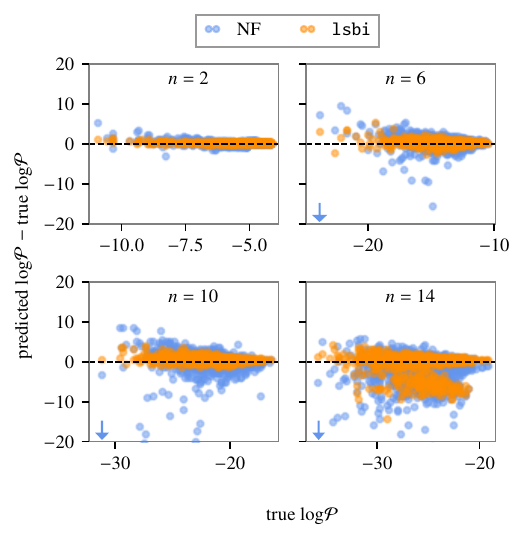}
    \caption{We evaluate the performance of normalizing flows on a mixture model, comprised of five Gaussians combined with unequal weights, as the number of dimensions increases. We generate samples from the mixture model in the full 14 dimensions using the package \textsc{lsbi}~\citep{lsbi_paper, lsbi_github} and drop the required number of columns to get samples in lower dimensions. We then train a normalizing flow using \textsc{margarine} on each set of samples, and compare the true log probability with the log probability predicted by the NF (blue). The black dashed line shows where the points would sit if the two perfectly matched. We also fit a five component Gaussian mixture model to each set of samples using \textsc{lsbi} and plot the log probability predictions of this too (orange). Since this model is in theory capable of fitting the distribution exactly, it could be taken to represent an upper bound on how well the task of density estimation can be performed in practice on this example. In lower dimensions, the NF performs well, albeit with slightly more scatter compared to the \textsc{lsbi} result. By $n=10$, however, the NF exhibits a significant decline in performance compared to the \textsc{lsbi} fit, with the most severe deterioration in the tails of the distribution. By $n=14$, both fits perform poorly. The arrows represent that there are points which lie outside the plot area. The full code to reproduce this plot, including details of how the mixture model was generated, can be found at~\protect\cite{zenodo}.}
    \label{fig:NF_vs_lsbi}
\end{figure}

We have found empirically that in many cases this method provides significant speedups compared with normal NS, with results that are in excellent agreement with the latter. Occasionally, however, the NF will learn a distribution which is narrower than the target `true' posterior. In these instances, sampling from the NF can become very inefficient and, in extreme cases, may provide biased results. This is because the peaks of the repartitioned likelihood can lie `deep' in the tails of the repartitioned prior. Even in more typical cases, the amount of acceleration provided by this method depends heavily on how well the flow has learned the posterior distribution provided by the low resolution pass of NS. For the number of dimensions that are involved in most gravitational wave problems, NFs can perform poorly at this density estimation task, especially in the tails of the distribution (see Figure~\ref{fig:NF_vs_lsbi}). This can severely limit the acceleration produced by this method for many realistic GW use cases. 

In this paper, we attempt to address these issues by replacing classic normalizing flows with what we christen $\beta$-flows. 

\subsection{$\beta$-flows and the connection with statistical mechanics}\label{statmech}

\begin{figure}
    \centering
    \includegraphics{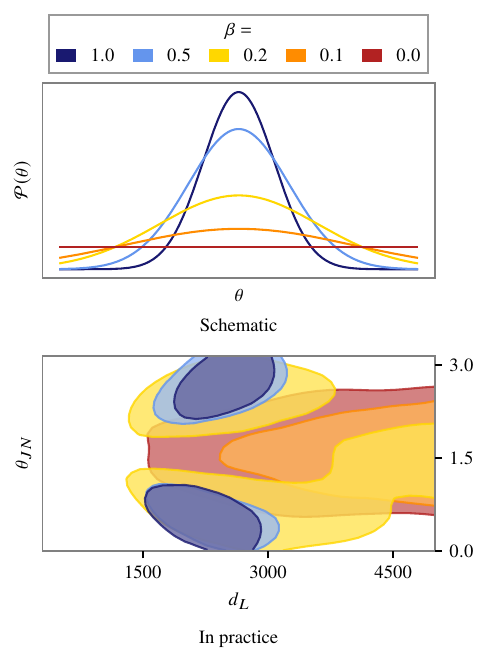}
    \caption{Nested sampling can emulate any temperature. The posterior has an inverse temperature of $\beta=1$ and the prior has an inverse temperature of $\beta=0$. In-between temperatures represent intermediate distributions. This is illustrated first on a more straightforward case where the posterior is a Gaussian and the prior is uniform (top panel). As $\beta$ decreases from $1$ to $0$, the distribution widens. The bottom panel shows the two-dimensional $1\sigma$ contours recovered from a simulated binary black hole merger for the luminosity distance, $d_L$, and the zenith angle between the total angular momentum and the line of sight, $\theta_\mathrm{JN}$. The posterior samples are re-weighted according to equation~\ref{beta_weights} to generate the distributions at various temperatures. Between $\beta=0.1$ and $\beta=0.2$, the distribution begins to split into two modes; in the statistical mechanics analogy, this is akin to a phase transition at the critical temperature. }
    \label{fig:temperature}
\end{figure}

There is an analogy to be made between the nested sampling algorithm and statistical mechanics~\citep{statmech_analogy}. In particular, the Bayesian evidence may be related to the partition function, if we consider the parameters $\theta$ to describe the microstate of a system with potential energy equal to the negative log-likelihood. The density of states may be expressed as:

\begin{equation}
    g(E) = \int \delta[E - E(\theta)] \pi(\theta) d\theta,
\end{equation}
where the prior is interpreted as the distribution of all possible states. An isolikelihood contour at $\mathcal{L}^\ast$ then corresponds to an energy limit $\epsilon = - \textrm{log}(\mathcal{L}^\ast)$. We can then see that the fractional prior volume, $X$, is simply the cumulative density of states, as a function of energy, rather than likelihood:

\begin{equation}
    X(\epsilon) = \int_{E(\theta) < \epsilon} \pi(\theta)d\theta = \int_{-\infty}^\epsilon g(E)dE. 
\end{equation}
The partition function at inverse canonical temperature $\beta$ may be rewritten as:

\begin{align}
    Z(\beta) & = \int e^{-\beta E}g(E) dE \notag = \int e^{-\beta \times - \textrm{log}\mathcal{L}(\theta)} \pi(\theta) d\theta \notag \\ 
    & = \int \mathcal{L}(\theta)^\beta \pi(\theta)d\theta = \int \mathcal{L}(X)^\beta dX  \label{eq:partition}
\end{align}
This inverse temperature ranges from $\beta = 0$, corresponding to an integral over the prior, to $\beta = 1$, recovering the Bayesian evidence integral from equation~\ref{evidenceX}. Though nested sampling is not thermal, it can simulate any temperature~\citep{Skilling2006NS}, meaning the partition function may be evaluated at any $\beta$ after the run (Figure~\ref{fig:temperature}). 

Generating samples at any inverse temperature involves modifying the posterior weights of the dead points from equation~\ref{dead_posteriorweight} to
\begin{equation}\label{beta_weights}
    p_i(\beta) = \frac{w_i \mathcal{L}_i^\beta}{\mathcal{Z}(\beta)}.
\end{equation}
$\mathcal{Z}(\beta)$ is evaluated from equation~\ref{eq:partition}. This functionality is provided by the package \textsc{anesthetic}~\citep{anesthetic}.
Typically, normalizing flows (NF) are trained only on the posterior samples, drawn from the $\beta=1$ distribution. As such, any information about the posterior and underlying likelihood functions encapsulated in the $\beta < 1$ intermediate distributions are discarded. The idea of $\beta$-flows is to incorporate this additional tail information to better learn the posterior.

\subsection{Training $\beta$-flows}

The goal is to learn a target distribution $\mathcal{P}(\theta)$ conditioned on the inverse temperature $\beta$ for samples from a NS run. We use conditional normalizing flows to transform samples from the multivariate base distribution $z \sim \mathcal{N}(0,1)$ onto $\mathcal{P}(\theta | \beta)$, where $\theta$ are drawn from the low resolution nested sampling run, with weights given by equation~\ref{beta_weights}. For any bijective transformation $f_\phi$, we can calculate the probability of a set of samples given $\beta$ by

\begin{equation}
    P_\phi(\theta|\beta) = \mathcal{N}(f_\phi(\theta,\beta)|\mu=0, \sigma=1) \left|\frac{df_\phi(\theta,\beta)}{d\theta}\right|.
\end{equation}
$\phi$ are the parameters of the neural network. We parameterize $f_\phi$ as a conditional masked auto-regressive (MAF) flow and train on a weighted reverse KL divergence~\citep{Bevins2023margarine2, Alsing2022anyprior}:

\begin{equation}
    \mathbb{L} = - \frac{1}{\sum p_i} \sum p_i(\beta) \textrm{log}P_\phi (\theta|\beta).
\end{equation}
We give the network samples weighted by various sets of $p(\beta)$, where $\beta$ ranges from $0$ to $1$. The training data therefore consists of $\{\theta, p(\beta), \beta\}$, in contrast to normal NFs, where we train with $\{\theta, p(\beta=1)\}$.

As $\beta$ increases from $0$ to $1$, the KL divergence between the weighted dead points and the prior increases non-linearly. The maximum KL divergence occurs at $\beta=1$, but the most rapid change happens at low $\beta$. 
\begin{equation}\label{DKL}
    \mathcal{D}_{\textrm{KL}} = \frac{1}{\sum_i p_i(\beta)} \sum_i p_i(\beta) \textrm{log} \frac{P(\theta | \beta)}{\pi(\theta)}.
\end{equation}
As such, instead of building the training data from $\beta$ values drawn uniformly from $[0,1]$, we define a $\beta$ schedule such that the change in KL divergence between subsequent sets of weighted dead points is constant. We choose a fixed number of $\beta$ values we want to train on first, and then calculate the exact $\beta$s between $0$ and $1$ that give equally spaced KL divergences.  


Once a $\beta$-flow has been trained on the samples from the low resolution first pass of NS, we then use this as a proposal for the high resolution pass. The flow can emulate not only the $\beta=1$ posterior, but also the intermediate distributions at any $0 \le \beta \le 1$. We treat $\beta$ as a hyperparameter, similar to the approach in~\cite{PR2} (though $\beta$ has a different meaning here), and sample over it during the high resolution run. Therefore, if the $\beta=1$ distribution is too narrow compared to the `true' posterior, the proposal can widen itself adaptively at runtime. The repartitioned prior and likelihood functions become
\begin{align}\label{PR_prior_beta}
\pi^\ast & = P(\theta | \beta) \\ 
\mathcal{L}^\ast & = \frac{\mathcal{L}(\theta)\pi(\theta)}{P(\theta | \beta)}\label{PR_likelihood_beta},
\end{align}
where this time the repartitioned prior and likelihood depend on $\beta$ (though the final evidences and posteriors will not).

\section{Results and discussion}\label{results}

In the following section, we present the results of applying the methods described above applied to both a simulated black hole binary (BBH) signal and a real event from the third Gravitational-Wave Transient Catalogue (GWTC-3). For each analysis, we first perform a low resolution pass of NS using \textsc{bilby}~\citep{bilby_paper}, with a slightly modified version of the built-in \textsc{PolyChord} sampler~\citep{Polychord1, Polychord2}. Specifically, the termination criterion in \textsc{PolyChord} is altered to be framed directly in terms of the change in the total estimated evidence, rather than the fraction of evidence remaining in the live points. For normal nested sampling, this alternative termination condition results in a very similar end point to the original. For further details on why and how this stopping criterion is changed, see Appendix~\ref{sec:appendix}.

Next, we train both a standard normalizing flow using \textsc{margarine}~\citep{Bevins2022margarine1, Bevins2023margarine2} and a $\beta$-flow, with code adapted from \textsc{margarine}, on the weighted posterior samples. Each of these trained flows respectively are then used as the repartitioned prior in a second pass of NS, where the likelihood is also repartitioned according to equation~\ref{PR_likelihood}. In this second pass, we use the same number of live points as in the first pass to facilitate a direct comparison between methods. However, in typical applications, a higher resolution pass would be used at this stage. All runs employ the \texttt{IMRPhenomXPHM} waveform model~\citep{IMRPhenomXPHM} and, unless otherwise specified, the standard BBH priors implemented in \textsc{bilby}. Plots are generated using \textsc{anesthetic}~\citep{anesthetic}.

\subsection{Injections}\label{sec:simulated}

We first demonstrate the method on a simulated BBH merger injected into Gaussian noise. We assume a two-detector configuration, with Hanford (H1) and Livingston (L1), and analyse $4$s of data. The signal is injected with the \texttt{IMRPhenomXPHM} waveform model, and the noise realization is set using the advanced LIGO O4 sensitivity curves. The binary has chirp mass $\mathcal{M}=28$\(M_\odot\) and mass ratio $q=0.8$. The spins are non-aligned, with an effective spin parameter $\chi_{\textrm{eff}}=0.27$ and it is located at a luminosity distance $d_L = 2000$ Mpc. The rest of the injected parameters are given in Table~\ref{tab:injected_params}. The network matched-filter signal-to-noise ratio (SNR) is $\rho_{mf} = 14.8$ and we show the posterior distributions obtained from a standard nested sampling run in Figures~\ref{fig:intrinsic_simulated} and~\ref{fig:eextrinsic_simulated}. Full posteriors are given in Appendix~\ref{sec:appendixB}.

\begin{table}
    \centering
    \begin{tabular}{|c|c|}
    \hline
    Parameter & Injected value \\
    \hline
    $\mathcal{M} / M_\odot$  & $28$ \\
    $q$    & $0.8$ \\ 
    $a_1$ & $0.4$ \\
    $a_2$ & $0.3$ \\
    $\theta_1$, rad & $0.5$ \\
    $\theta_2$, rad & $1.0$ \\
    $\phi_{12}$, rad & $1.7$  \\
    $\phi_{JL}$, rad & $0.3$  \\
    $d_L$, Mpc & $2000$ \\
    $\theta_{JN}$, rad & $0.4$  \\
    $\psi$, rad & $2.66$  \\
    $\phi$, rad & $1.3$  \\
    $\alpha$, rad & $1.375$  \\
    $\delta$, rad & $-1.21$ \\
    $t_c$, GPS time & $1126259642.413$ \\
    \hline
    \end{tabular}
    \caption{The injected parameters for the simulated BBH signal are shown. For a definition of the parameters, see Table E1 of~\protect\cite{Bilby_parameters}.}
    \label{tab:injected_params}
\end{table}

For the first step of our method, we perform a low resolution NS run with $n_{\textrm{live}}=200$; this is a much lower number of live points than what is typically used in standard 15-parameter gravitational wave analyses, but is still high resolution enough to capture the main features and modes of the posterior. We then use the weighted samples from this to train both a NF and a $\beta$-flow. It is important to note that it is possible for the low resolution run to miss small secondary modes and features of the true posterior, leading to issues with PR if this is then used as the repartitioned prior. This is one of the main benefits of using $\beta$-flows, and is discussed further in Section~\ref{sec:realdata}. The relative performances are shown in Figure~\ref{fig:NF_betaflow_simulated}, where the predicted probabilities from the flows are compared to the posterior probabilities given by NS. Both flows exhibit a fairly large scatter about the target probabilities, typical for a 15-dimensional problem, but the $\beta$-flow performs noticeably better than the NF, particularly in the tails of the distribution. 

Each flow is then used as the updated prior for a PR NS run, also with $n_{\textrm{live}}=200$, and the evidences and posteriors obtained from this run are compared to those from standard NS analyses with the same number of live points. Figure~\ref{fig:logZ_simulated} shows the log evidence distributions obtained from each PR run and from the original low resolution pass of NS. The results are in excellent agreement, with the error bars on $\log\mathcal{Z}$ being tighter for both the PR runs compared to normal NS, despite using the same number of live points, as predicted by equation~\ref{uncertainty}. We also compare the posteriors obtained from each method, which are plotted in Figures~\ref{fig:intrinsic_simulated} and~\ref{fig:eextrinsic_simulated} and again show good agreement between the methods. 

Table~\ref{tab:simulated_speedup} outlines the relative acceleration provided by each flow compared to normal NS. For a fixed uncertainty in $\mathrm{log}\mathcal{Z}$, given that $n_\textrm{live} \propto D_\mathrm{KL}$, we may rewrite equation~\ref{runtime} as

\begin{equation}
    T \propto T_\mathcal{L} \times f_\textrm{sampler} \times \mathcal{D}_{\textrm{KL}}^2.
\end{equation}
Then, the precision-normalized acceleration of the PR run may be approximated as
\begin{equation}
    \frac{T^{\textrm{normal NS}}}{T^{\textrm{PR NS}}} = \left( \frac{\mathcal{D}_{\mathrm{KL}}^{\textrm{normal NS}}}{\mathcal{D}_\mathrm{KL}^{\textrm{PR NS}}} \right) ^2
\end{equation}
Using PR in conjunction with a trained $\beta$-flow led to almost an order of magnitude improvement in the runtime (see Figure~\ref{fig:truncation}). In this instance, the NF performs similarly well to the $\beta$-flow, indicating that the NF has learned a wide enough distribution to avoid sampling inefficiencies in the PR run. 

It is important to note at this stage that the quoted speedup factors are calculated purely based on the number of iterations that would be required for a precision-normalized PR run. It does not take into account the changes to $T_\mathcal{L}$, the time for a single likelihood evaluation, from including the flows in the likelihood. The $\beta$-flow took longer to evaluate than the NF we used. This also means that for analyses using a waveform model like \texttt{IMRPhenomXPHM}, $T_\mathcal{L}$ increases by such a factor that we do not recommend using $\beta$-flows in their current form in these cases. This point is addressed further in the conclusions, including a discussion of future work to speed up the evaluation of our $\beta$-flows, but for now, we intend for the methods presented in this paper to be used in analyses where the evaluation of the gravitational wave likelihood is of comparable cost to the evaluation of the $\beta$-flow. We also note that, strictly speaking, the speedup factors should include the time it takes to perform the original low resolution NS run, but in the typical case where the second pass of NS uses a much larger number of live points, this cost will not contribute significantly to the overall runtime.

\begin{figure}
    \centering
    \includegraphics{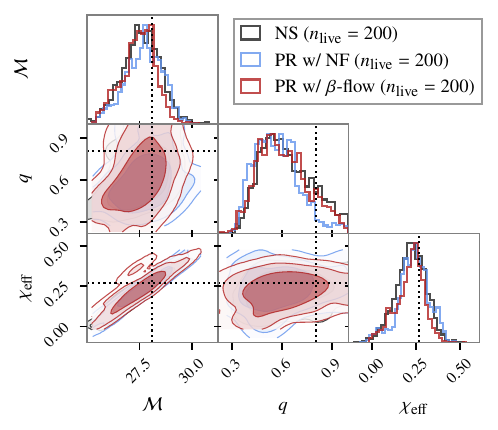}
    \caption{The posteriors obtained on some intrinsic parameters (chirp mass $\mathcal{M}$, mass ratio $q$ and effective spin parameter $\chi_\mathrm{eff}$) from standard NS are compared to those obtained using PR with normalizing flows or $\beta$-flows. The results are consistent, showing both the PR methods have managed to recover the same answers as normal NS.}
    \label{fig:intrinsic_simulated}
\end{figure}

\begin{figure}
    \centering
    \includegraphics{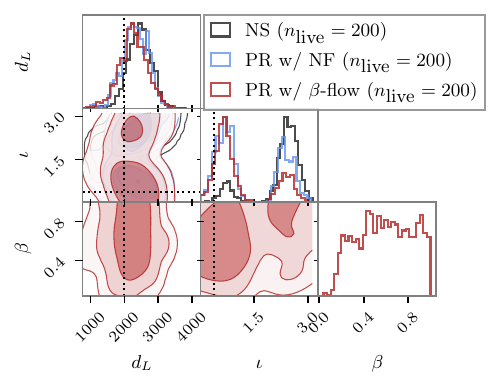}
    \caption{Similarly to~\ref{fig:intrinsic_simulated}, the posteriors on the extrinsic parameters, the luminosity distance and inclination, from the two methods are compared. Again, the results are comparable, with the PR NS methods able to achieve this with far fewer likelihood evaluations. The $\beta$-flow method gives less posterior weight in the second mode and more posterior weight in the first mode than the normal NS run, but this could occur from two separate normal NS runs too, due to the stochasticity of NS~\citep{Adam, Polychord1}. This stochasticity is quantified by the $\text{log}\mathcal{Z}$ error bars that \textsc{PolyChord} outputs for individual clusters.}
    \label{fig:eextrinsic_simulated}
\end{figure}

\begin{figure}
    \centering
    \includegraphics{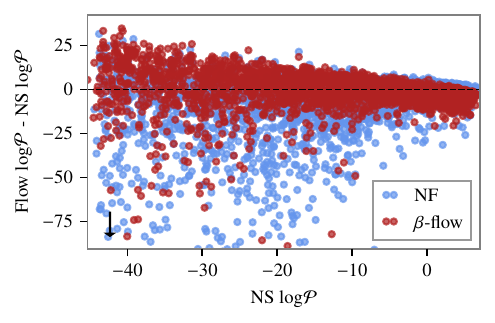}
    \caption{We compare how well both the typical normalizing flow (NF) and the $\beta$-flow (evaluated at $\beta=1$) have learned the rough posterior from the low resolution pass of NS. If the flows have learned the posterior perfectly, the points should lie on the black dashed line. The arrow indicates that there are points which lie below the axes. The $\beta$-flow predictions display much less scatter about this line, showing that the extra tail information from the NS temperature has indeed enabled the flow to learn the posterior better. Although the scatter on the NF seems large, this is an empirically typical performance on a 15-dimensional problem.}
    \label{fig:NF_betaflow_simulated}
\end{figure}

\begin{figure}
    \centering
    \includegraphics{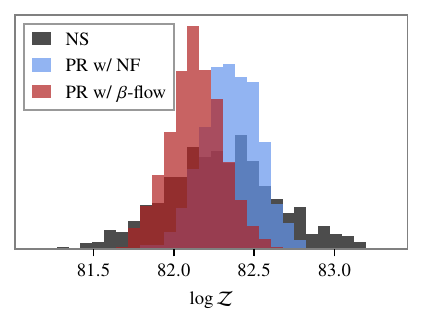}
    \caption{The $\log\mathcal{Z}$ estimates calculated using \textsc{anesthetic} for normal NS, posterior repartitioned NS with a normalizing flow, and posterior-repartitioned NS with a $\beta$-flow are compared. All of the runs are performed with $n_{\textrm{live}}=200$ for easier comparison. The estimates are all consistent with each other, but both the PR runs have smaller error bars, as expected. }
    \label{fig:logZ_simulated}
\end{figure}

\begin{figure}
    \centering
    \includegraphics{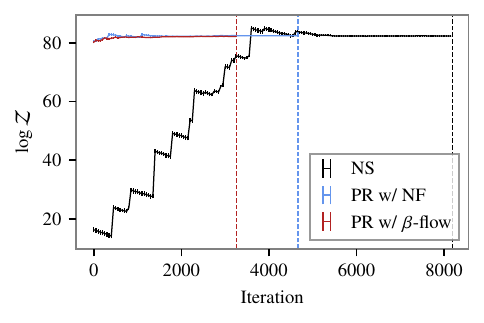}
    \caption{During a normal NS run, the evidence is accumulated as the live points compress towards the peak of the likelihood. The total evidence estimate for normal NS (black) becomes stable late into the run, only after the live points occupy a very small fraction of the prior volume. Because the updated prior for the posterior repartitioned runs is roughly the posterior from the low resolution pass of NS, most of the evidence has already been accumulated very early on in the run. We keep running until the total evidence estimates for the accelerated runs (blue and red) have stabilized. This happens much earlier than for normal NS, and the live points typically still occupy a significant fraction of the prior volume.}
    \label{fig:truncation}
\end{figure}

\begin{table}
    \centering
    \begin{tabular}{|c|c|c|c|c|c|}
    \hline
     type  & $n_\textrm{live}$ & $N_\textrm{iter}$ & $ln(\mathcal{Z})$ & speedup \\
    \hline
     normal NS    & 200 & 8186 & $82.28 \pm 0.35$ & - \\
     PR NS w/ NF & 200 & 4663 & $82.37 \pm 0.18$ & $\times 7$\\
     PR NS w/ $\beta$-flow& 200 & 3252 & $82.14 \pm 0.18$& $\times 9$ \\
     \hline
    \end{tabular}
    \caption{For the simulated event, results of the runs comparing normal NS to posterior-repartitioned NS (PR NS) are shown. $N_iter$ is the total number of iterations, $i$, of the algorithm that were performed, and is proportional to the number of likelihood evaluations. Both the run using a typical normalizing flow and using a $\beta$-flow finish significantly sooner than normal NS. The final column shows the \textbf{precision-normalized} speedup, calculated by using equation~\ref{uncertainty} to work out how many live points we would need to run with in order to match the $\log\mathcal{Z}$ uncertainty of the normal NS run, and then scaling $N_{\textrm{iter}}$ proportionally.}
    \label{tab:simulated_speedup}
\end{table}

\subsection{Real Data}\label{sec:realdata}

We demonstrate the above methods on the real event, GW191222\_033537 (henceforth GW191222) from GWTC-3, chosen in part due to the multi-modality and complex shape of its posteriors, to illustrate the effects of this on PR. GW191222 was a two detector event, with a network match-filtered SNR of $12.5$, and we analysed $8$s of data.

As before, we perform a low resolution pass of NS on which we train both flows. This time, however, we use $350$ live points. The posterior for this event is more complex and has more multi-modality than the simulated example above, so we give the flows more samples to train on in order to give them a better chance of learning these features accurately. We do not include any additional parameters in our analysis to account for uncertainty in the calibration of detectors, meaning that, as in Section~\ref{sec:simulated}, we are sampling over $15$ parameters. Inclusion of these additional calibration parameters is left for further work.

\begin{figure}
    \centering
    \includegraphics{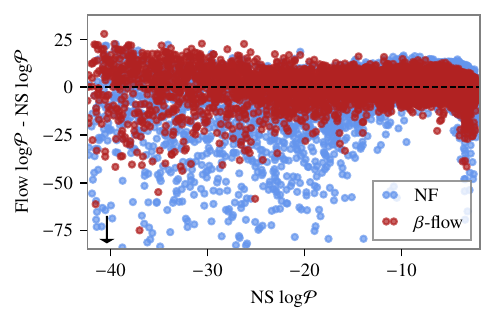}
    \caption{The $\beta$-flow once again performs better at predicting the log probability given by NS. This time, both flows have a larger spread at higher log probabilities and a `tail' of points below the black dashed line. Again, the arrow indicates that there are points which lie below the axes. The NF heavily under-predicts the posterior probability of certain samples, which is indicative of the fact that it has failed to capture the multi-modality of the rough posterior. }
    \label{fig:GW191222_NF_v_betaflow}
\end{figure}

As shown in Figure~\ref{fig:GW191222_NF_v_betaflow}, once again the $\beta$-flow is able to learn the rough posterior from the NS run more accurately, and is better at predicting deep tail probabilities than the NF. However, both flows exhibit a wider spread than before at the highest log probability values, and there is a `tail' of under-predictions for certain samples from the peak of the posterior. This is indicative of the fact that the full multi-modality of the NS posterior has not been captured by either flow, though the NF does perform significantly worse. This is key to understanding the final results.

\begin{figure}
    \centering
    \includegraphics{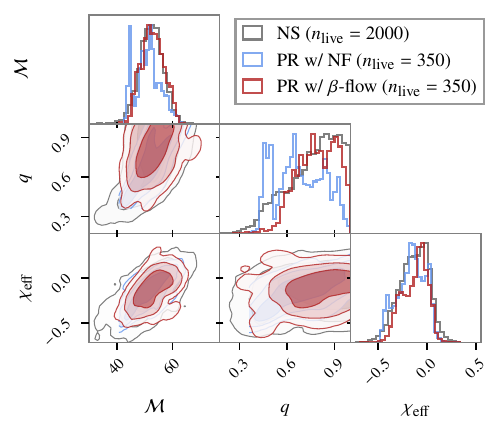}
    \caption{Unlike for the previous simulated signal, using the trained NF as the proposal for PR NS has led to biased results. The NF learned the posterior from the low resolution run poorly, and without the ability to widen itself at runtime, this has produced incorrect posteriors and evidence. The $\beta$-flow is robust to this issue as the proposal is over all values of $\beta$. This means that even if the learned flow is too narrow or has not learned the multi-modality sufficiently well, it can still adapt the proposal at runtime and, in the worst case scenario, samples will simply be drawn from the original prior ($\beta$=0).}
    \label{fig:GW191222_intrinsic}
\end{figure}

\begin{figure}
    \centering
    \includegraphics{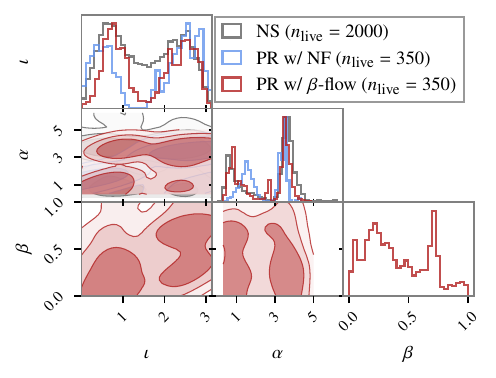}
    \caption{The multi-modality in the extrinsic parameters has caused a biasing effect for PR NS with the NF, since the NF did not learn all modes properly. The posterior on $\beta$, the inverse temperature, for the $\beta$-flow run is also included. If the flow learned the rough posterior well, we would expect to see a uniform posterior on $\beta$. The low posterior probability at $\beta=1$ indicates that the $\beta$-flow had to widen itself at runtime due to the $\beta=1$ distribution being unsuitable as a prior.}
    \label{fig:GW191222_extrinsic}
\end{figure}

To properly verify whether we have recovered the correct posteriors for this real event, we compare our posteriors from the accelerated methods to those from a higher resolution ($n_\mathrm{live} = 2000$) standard NS run. Since the NF does not learn the multi-modality of the posterior well enough, it sets the proposal for the PR run such that certain modes are only included in the prior with very low probabilities. This leads to a biasing of the final posteriors, shown in Figures~\ref{fig:GW191222_intrinsic} and~\ref{fig:GW191222_extrinsic}. The $\beta$-flow also doesn't fully learn the multi-modality of the posterior, but since it acts as an adaptive prior at runtime, able to draw samples from the distribution at any inverse temperature, it does not completely cut off important regions of the parameter space in the same way the NF does. This property also makes PR with $\beta$-flows more robust to cases where the low resolution nested sampling run has missed secondary modes and features. Looking at the posteriors in Figure~\ref{fig:GW191222_extrinsic}, we can indeed see that the $\beta=1$ distribution was too narrow and excluded regions of the parameter space with non-negligible posterior weight. Otherwise, we would expect to see a roughly uniform posterior on $\beta$, but instead we see that $\beta=1$ has a low posterior probability. 

\begin{figure}
    \centering
    \includegraphics{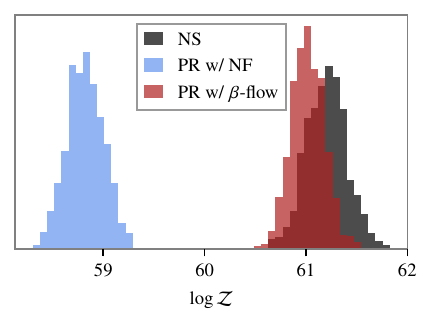}
    \caption{$\log\mathcal{Z}$ estimates calculated using \textsc{anesthetic} are compared. The NF learned the rough posterior from the low resolution run poorly, insufficiently capturing its multi-modality. This has led to a biasing of the final evidences and posteriors, since the proposal from the NF cannot widen itself like the $\beta$-flow can. The $\beta$-flow not only learned the distribution from the first pass of NS better, but also enabled an adaptive proposal at runtime, ensuring robustness against such biases. }
    \label{fig:GW191222_logZ}
\end{figure}

The evidence calculated by PR NS using the NF also reflects this bias (Figure~\ref{fig:GW191222_logZ}). The results are incompatible with those from normal NS, and is another sign that regions of the parameter space with significant posterior weight were missed due to the updated prior being too narrow. Once again, because the $\beta$-flow can emulate any temperature, it is more robust to these issues and is able to give better results than the NF, despite a poor performance at the posterior density estimation.

As for the consistency of the posteriors, Figures~\ref{fig:GW191222_intrinsic} and~\ref{fig:GW191222_extrinsic} generally show agreement with the standard NS results, but certain parameters, such as the mass ratio and inclination, exhibit some differences. In order to validate the results further, we performed a high resolution PR run with the $\beta$-flow, the full posteriors from which are presented and discussed in Appendix~\ref{sec:appendixC}. The differences in the mass ratio no longer appear, but there are some differences in other parameters, particularly in those that are not well constrained. This is possibly due to the stochasticity associated with sampling a heavily multi-modal posterior, and this is explored further in the Appendix. 

\begin{figure}
    \centering
    \includegraphics{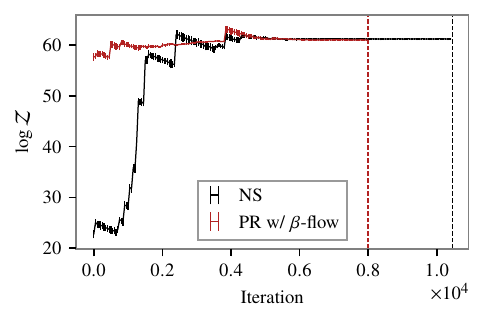}
    \caption{PR NS with the $\beta$-flow terminates before normal NS with the same number of live points. The precision-normalized speedup is less than for the simulated example, but is still a factor of two faster. We don't show the equivalent line for the NF because it failed to correctly recover the evidence and posterior.}
    \label{fig:GW191222_truncation}
\end{figure}

Since the $\beta$-flow did not learn the posterior at $\beta=1$ as well as for the simulated case, the speedup given by using this flow as the updated prior was not as large (Figure~\ref{fig:GW191222_truncation}). The exact acceleration provided by PR NS is very sensitive to the accuracy of the density estimation. However, the precision-normalized runtime was still twice as fast as for normal NS and, importantly, we demonstrate the robustness of this method in giving reliable evidences, even when the density estimation is relatively poor quality. The worst case scenario of using PR NS with $\beta$-flows is that we get correct evidences which take the same amount of time as normal NS (since for a very poor $\beta$-flow we would sample preferentially from the $\beta=0$ distribution, which is the original NS prior). The same cannot be said for PR NS with NFs, however, and the results in this section give an example where this method breaks down completely. For this reason, we recommend using $\beta$-flows in place of NFs when implementing posterior repartitioning. 

\begin{table}
    \centering
    \begin{tabular}{|c|c|c|c|c|c|}
    \hline
     type  & $n_\textrm{live}$ & $N_\textrm{iter}$ & $ln(\mathcal{Z})$ & speedup\\
    \hline
     normal NS    & 350 & 10445 & $61.21 \pm 0.21$ & - \\
     PR NS w/ $\beta$-flow& 350 & 7995 & $61.02 \pm 0.17$ & $\times 2$ \\
     \hline
    \end{tabular}
    \caption{Normal NS is compared to the PR NS method for real event GW191222. PR NS with the $\beta$-flow is twice as fast as normal NS for a \textbf{precision-normalized} run. This is a smaller speedup than for the simulated example, and this is driven by the fact that the $\beta$-flow was not able to learn the rough posterior from pass 1 as accurately. PR NS with the NF is not shown here; although it was also quicker than normal NS, it gave incorrect posteriors and evidences due to the biased proposal.}
    \label{tab:GW191222_speedup}
\end{table}

\section{Conclusions}\label{conclusions}

In this paper, we outline how posterior repartitioning using normalizing flows can accelerate nested sampling. While we demonstrate these methods with \textsc{PolyChord}, this is a general acceleration technique applicable to a variety of nested sampling algorithms, and does not inherently rely on machine learning to be effective. Bringing together previous work~\citep{PR1, PR2, Supernest, Bevins2022margarine1, Bevins2023margarine2, Alsing2022anyprior}, we demonstrate this method on realistic gravitational wave examples. However, there are a few drawbacks of using traditional normalizing flow architectures in posterior repartitioned nested sampling. Firstly, the amount of acceleration provided by PR NS is highly dependent on the success of the flow in learning the posterior distribution provided by the low resolution nested sampling run. In particular, the more successful the flow is at learning the deep tail probabilities, the sooner we can terminate the high resolution PR run. However, we empirically show that the accuracy of commonly used NF architectures is often poor in the tails of the target distribution, especially as the dimensionality increases. Furthermore, if the distribution learned by the flow is too narrow compared to the true posterior, this can lead to sampling inefficiencies, making the problem harder, and in the worst case scenario can give biased results. We show a real GW case where this occurs.

In order to mitigate these issues, we introduce $\beta$-flows, which are conditional normalizing flows trained on nested samples and conditioned on inverse temperature, $\beta$.
$\beta$-flows are shown to be better at predicting deep tail probabilities than traditional normalizing flows, as they have access to intermediate distributions between prior and posterior during training, as opposed to just the posterior samples. Additionally, $\beta$-flows can emulate not just the target posterior distribution itself, which corresponds to $\beta=1$, but also any of these intermediate distributions. At runtime, we sample over different values of $\beta$, meaning that if the $\beta=1$ distribution learned by the flow is indeed too narrow, the repartitioned prior can adaptively widen itself at runtime to mitigate sampling inefficiencies and biases. For the same case on which normal normalizing flows fail, we show that replacing normalizing flows with $\beta$-flows results in much more consistent posteriors and evidences, though they still exhibit some differences from standard NS in certain parameters, particularly unconstrained ones.

One current disadvantage of $\beta$-flows is that, due to the flow having to store and call more biases and weights, they take significantly longer to evaluate than more typical normalizing flows. For evaluating the probability of a single sample, they take about $100$ms, $100$ times slower than the NF trained using \textsc{margarine}. This limitation could be ameliorated in a few ways. Firstly, the $\beta$-flow could be implemented in \textsc{jax}, which could significantly reduce this cost, though it would likely still be more expensive than a standard NF. Moreover, NFs and $\beta$-flows are designed to evaluate batches of samples at once, and so this cost does scale linearly with the number of samples. For a set of $10,000$ samples, the $\beta$-flows only take twice as long to evaluate them as for a single sample, and only take $4$ times as long as a flow using \textsc{margarine}. Therefore, if we could implement PR within a nested sampling algorithm which can properly make use of this property of normalizing flows, the cost to evaluate the $\beta$-flow would become negligible. Both of these are promising avenues for future work on this topic, and would make the methods presented in this paper suitable for a wider range of likelihoods. In their current form, they can still be worthwhile implementing in cases where the likelihood itself is of comparable computational cost to the flows. 

Currently, the method requires nested samples from the exact likelihood we want to use in our final analysis in order to train the flows. Future work could involve adapting the methodology to enable the $\beta$-flow to learn an approximate distribution, perhaps from a cheaper waveform model, and then use this as a proposal for the high resolution run. This has synergies with likelihood reweighting~\citep{likelihood_reweighting} and tempered importance sampling~\citep{tempered_likelihood_reweighting}. $\beta$-flows also have a connection with continuous normalizing flows (CNFs) and diffusion models, where there is a natural user tunable parameter akin to $\beta$~\citep{CNF}. Future work could explore this link, and could explore using CNFs in conjunction with posterior repartitioning too.

\section*{Acknowledgements}

MP was supported by the Harding Distinguished Postgraduate Scholars Programme (HDPSP). WH was supported by a Royal Society University Research Fellowship. HTJB acknowledges support from the Kavli Institute for Cosmology, Cambridge, the Kavli Foundation and of St Edmunds College, Cambridge.

This work was performed using the Cambridge Service for Data Driven Discovery (CSD3), part of which is operated by the University of Cambridge Research Computing on behalf of the STFC DiRAC HPC Facility (www.dirac.ac.uk). The DiRAC component of CSD3 was funded by BEIS capital funding via STFC capital grants ST/P002307/1 and ST/R002452/1 and STFC operations grant ST/R00689X/1. DiRAC is part of the National e-Infrastructure.

\section*{Data Availability}

All the data used in this analysis, including the relevant nested sampling dataframes, can be obtained from~\cite{zenodo}. We include a notebook with all the code to reproduce the plots in this paper. We also include an example~\textsc{python} file to show how to implement posterior repartitioning in~\textsc{bilby}, with instructions on how to modify the~\textsc{bilby} source code. The code we used for training the $\beta$-flows in this paper is publicly available and can be found at~\cite{betaflows_github}. The modified version of \textsc{PolyChord} used to perform these analyses can be found at~\cite{polychord_alt_terminate_github}.



\bibliographystyle{mnras}
\bibliography{example} 

\appendix

\section{Termination conditions for NS}\label{sec:appendix}
Posterior repartitioned NS has slightly different properties to normal NS. This means that the usual termination condition that is used for the latter is too cautious for the former. Nested sampling compresses live points exponentially towards the peak of the likelihood function. As they close in on the peak, the likelihood values begin to saturate ($\mathcal{L}_i \rightarrow \mathcal{L}_{\textrm{peak}}$) and the fractional volumes become very small ($X_i \rightarrow 0$)~\citep{Keeton_2011}. As such, beyond a certain point there are diminishing returns for performing further iterations of the algorithm.

\begin{figure}
    \centering
    \includegraphics{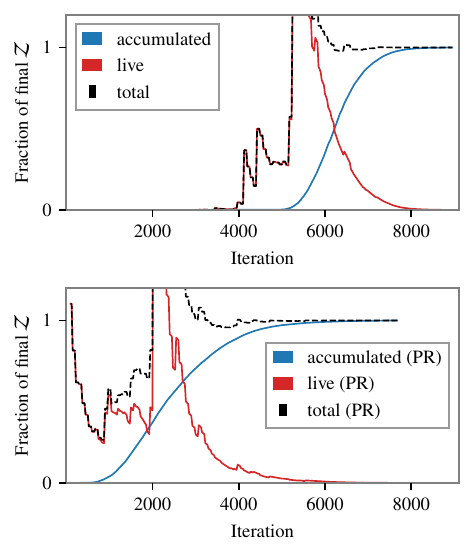}
    \caption{As described in~\protect\cite{Keeton_2011}, the total evidence estimate varies throughout a typical NS run. In typical NS (top panel), the accumulated evidence before we reach the bulk of the posterior is very low, due to small likelihood values. When the live points enter the posterior bulk, this accumulated evidence steadily increases, until the likelihoods saturate and the fractional prior volume changes become negligible. The estimate of the evidence remaining in the live points is much more unstable, and is initially dominated by a single live point with the highest weight, $w_i \mathcal{L}_i$. It spikes and falls rapidly as a new live point is found which temporarily dominates the live evidence, and hence the total evidence estimate also changes. This total evidence estimate usually only becomes stable once the fractional evidence remaining in the live points is small, making this a robust proxy for the stopping criterion in normal NS. When doing posterior repartitioning, however, the total evidence estimate may stabilize before the live evidence fraction has fallen by the required amount (bottom panel). In these cases, the algorithm may continue for many more iterations without any additional benefit. Here, the usual termination condition is too cautious and should be framed directly in terms of the total evidence estimate instead.}
    \label{fig:Keeton}
\end{figure}

At each iteration $k$, the estimated total evidence is the sum of the accumulated evidence and the estimated evidence remaining in the live points.

\begin{equation}
    \mathcal{Z}_\mathrm{tot} = \mathcal{Z}_\mathrm{dead} + \mathcal{Z}_\mathrm{live} \approx \sum_{i=1}^{k} \mathcal{L}_i(X_{i-1} - X_i) + \bar{\mathcal{L}}_\textrm{live} X_k.
\end{equation}
$\bar{\mathcal{L}}_\textrm{live}$ represents the average likelihood of the live points at iteration $k$, and $X_k$ is the remaining fractional volume.

Figure~\ref{fig:Keeton} shows the evolution of each of these terms as a function of the iteration number. Initially, since the deleted points have not yet reached the bulk of the posterior, the total accumulated evidence is very small due to low likelihoods. Once the bulk of the posterior is reached, the accumulated evidence builds up rapidly as the likelihood increases, until the likelihood flattens out near the peak and the fractional volume changes become negligible. At this point, the accumulated evidence saturates. 

The estimated live evidence is very unstable to begin with. It is usually dominated by a single live point which lies in the posterior, and rises sharply when a new live point is found which temporarily becomes the main contributor, falling again as the fractional prior volume decreases. Once the live points are completely contained within the bulk of the posterior, the estimated live evidence begins to fall smoothly, unless previously missed modes are found. The total evidence is also unstable at the beginning, dominated by the live evidence, but starts to become stable once we enter the posterior bulk. Ideally, we would terminate our run once this estimated total evidence has become completely stable and does not change significantly as we perform further iterations of the algorithm. 

\begin{figure}
    \centering
    \includegraphics[]{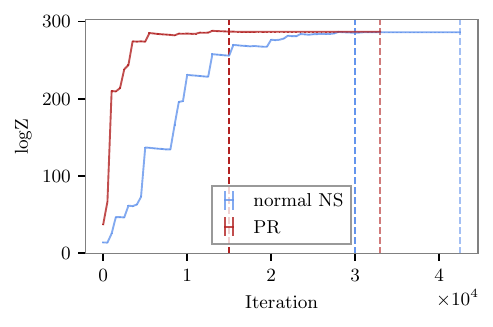}
    \caption{Example normal NS and PR NS runs were performed on simulated data, using the usual termination condition based on the live evidence fraction. Using post-processing tools in \textsc{anesthetic}, we can examine what the $\log$Z estimate would have been had we terminated the run earlier. \textsc{anesthetic} takes the dead points upto iteration $i$, and adds on the live points at iteration $i$, recalculating the weights accordingly, to give the total $\log$Z estimate if the run had been terminated at this iteration. For normal NS, we see that the $\log$Z estimate we obtain from the run would not have changed significantly after about iteration $i = 30,000$, but the run continues for a further $12,000$ iterations to wait for the live evidence fraction to become low enough. For PR NS, the $\log$Z estimate would have been the same had we terminated our run at iteration $i = 15,000$, but we continue to run the algorithm for another $18,000$ iterations to compress the live points enough. This shows that at the end of a PR NS run, the live points are compressing more slowly, but we have obtained a stable evidence estimate well before they compress to the required degree, meaning we are performing additional iterations for minimal gain.}
    \label{fig:appendix2}
\end{figure}

In most cases, a proxy for this is to stop when the estimated live evidence is some very small fraction of the total accumulated evidence, and this is the default termination condition in many popular NS implementations~\citep{Ashton2022NSReview}. In the specific case of posterior repartitioning, however, this is perhaps too cautious a stopping criterion. In the extreme case where our trained flow has perfectly learned the posterior distribution, we could terminate our high resolution PR run almost immediately, since although performing further iterations of the algorithm would increase the accumulated evidence and decrease the live evidence, it would make no difference to the total evidence estimate. Even in the case where the flow has imperfectly learned the posterior, much of the discrepancy is likely to be in the tails of the distribution (see e.g. Figure~\ref{fig:truncation}). As such, the total evidence estimate would still likely stabilize well before the live evidence fraction falls below the usual threshold. This is illustrated further in Figure~\ref{fig:appendix2}, where we show what happens to the $\log$Z estimate if the runs were terminated earlier than by the usual termination condition. As a result, in the above analyses we modified \textsc{PolyChord} to set the termination condition for the run in terms of the estimated total evidence directly, instead of the live evidence fraction. We set the new condition such that the run terminates when the total estimated evidence has not changed by more than $0.01$\% over the previous $5 \times n_\textrm{live}$ iterations. These values were chosen so that for normal NS, this results in a very similar end point to the default condition for all the examples we ran. 

\section{Simulated Data Full Posteriors}\label{sec:appendixB}

\begin{figure*}
    \centering
    \includegraphics[width=\linewidth]{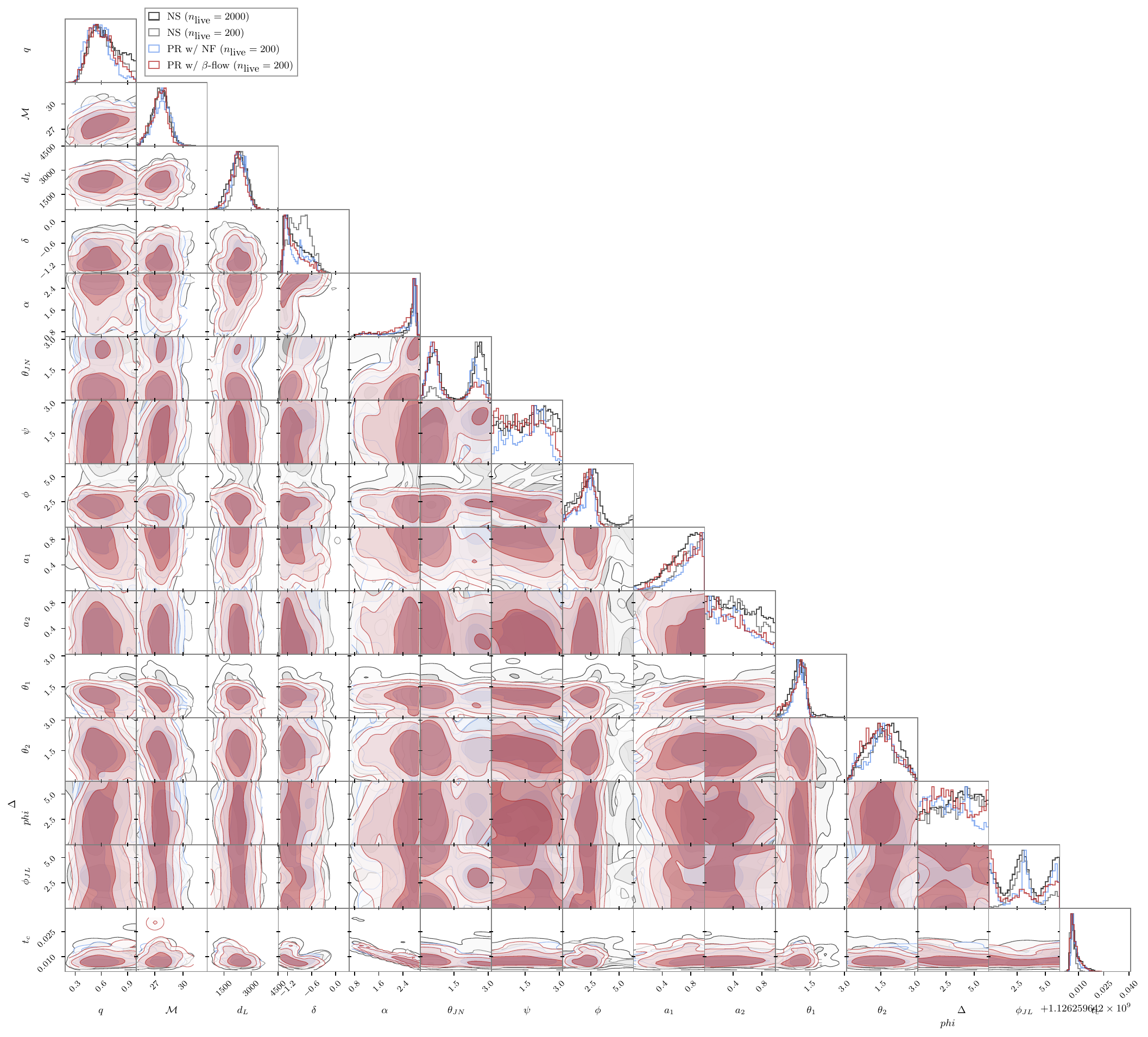}
    \caption{Full posteriors for the simulated example. The light grey and black show the low and high resolution standard NS runs respectively.}
    \label{fig:B1}
\end{figure*}

Figure~\ref{fig:B1} shows the full posterior distributions for the simulated example discussed above. We plot both the low resolution and high resolution nested sampling runs, and the low resolution PR runs. The PR run with the $\beta$-flow generally shows good agreement with the standard NS results. In parameters where the posterior is multi-modal, such as $\theta_\textrm{JN}$, the $\beta$-flow run shows less posterior weight in one of the modes than the standard NS runs, but this could occur from two separate normal NS runs too, due to the stochasticity of NS~\citep{Adam, Polychord1}. This stochasticity can in theory be quantified by the $\text{log}\mathcal{Z}$ error bars that \textsc{PolyChord} outputs for individual clusters, though these runs were performed with clustering turned off in order to more closely match standard GW analyses. The parameters in which the results are least consistent are the ones where the posteriors are not very well constrained. It is also important to note that in the phase parameter, neither the $\beta$-flow nor the NF PR runs are in agreement with the NS posteriors at larger phase values. This could be due to the flows struggling to learn the multi-modal phase distribution, but could also be due to the lack of a periodic boundary condition being implemented for this parameter.

\section{Real Data Full Posteriors}\label{sec:appendixC}

To further validate and understand the results from the $\beta$-flow, we performed a high resolution PR run with $2000$ live points to compare with the high resolution standard NS run. We also performed a second reference run of normal NS using the sampler \textsc{dynesty} to better understand the inherent stochasticity of sampling such a multi-modal posterior. We note that a second run performed with \textsc{PolyChord} also exhibited similar differences to the first \textsc{PolyChord} run as the \textsc{dynesty} run, but we do not show the results of the second \textsc{PolyChord} run so as not to overcrowd the plot. 

\begin{figure*}
    \centering
    \includegraphics[]{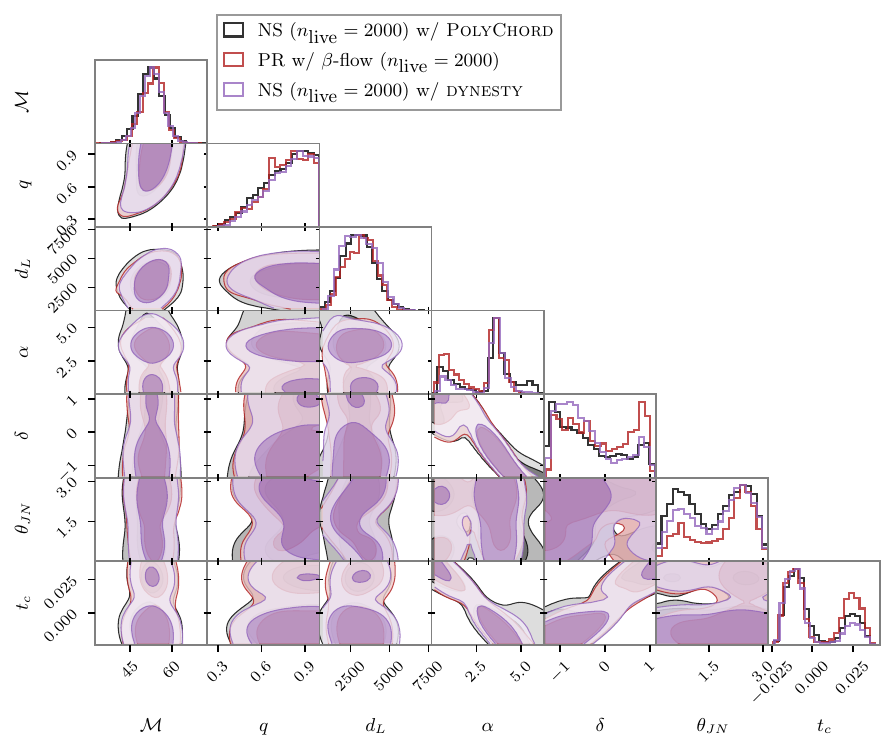}
    \caption{Posteriors on some of the parameters are shown. The high resolution $\beta$-flow run is mostly in agreement with the high resolution standard NS run, but there are some differences visible in the sky location and inclination parameters. It is likely that these stem from stochasticity of sampling such a multi-modal posterior. For reference, a normal nested sampling run performed using \textsc{dynesty} is also shown. This demonstrates the scale of differences that can be expected in the posteriors due to stochasticity.}
    \label{fig:C1}
\end{figure*}

\begin{figure*}
    \centering
    \includegraphics[]{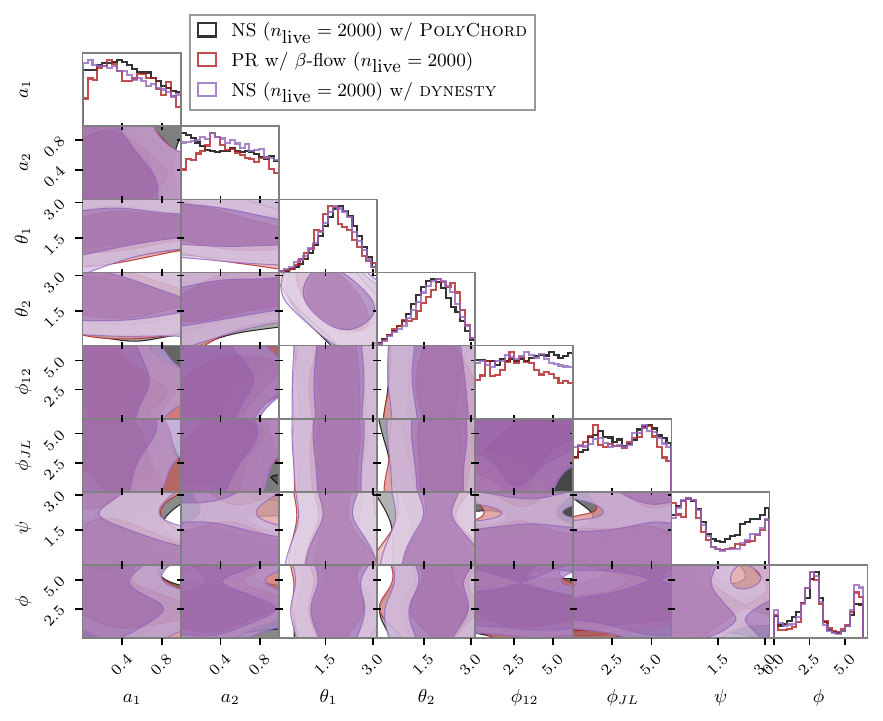}
    \caption{The rest of the parameters are shown. The PR NS run with the $\beta$-flow exhibits differences in the recovered posteriors on a few of the parameters. A second reference run performed with \textsc{dynesty} is also shown, and the differences between the PR NS and normal NS runs with \textsc{PolyChord} are of a similar order to the differences between the two normal NS runs performed with different samplers. It is likely that these difference arise due to the multi-modality of this posterior, increasing the stochasticity associated with sampling. \textsc{PolyChord}, when run in its clustering mode, can quantify this stochasticity through error bars on the evidences from individual clusters.}
    \label{fig:C2}
\end{figure*}

The full posteriors for all $15$ parameters are shown in Figures~\ref{fig:C1} and~\ref{fig:C2}. The results are generally in agreement, though there are a few differences to note. Firstly, although it appeared from Figure~\ref{fig:GW191222_intrinsic} that the mass ratio posteriors for the two runs were not entirely consistent, we see that for a higher resolution PR run, they are indeed in agreement. The main differences between the two posteriors are again in parameters that have not been well constrained. Beyond this, the $\beta$-flow run also exhibits from differences in the tilt angle parameters. Once again, we see that in multi-modal parameters like $\theta_\textrm{JN}$, the PR run shows a decreased posterior probability in one mode compared to the standard high resolution NS, but the two reference runs also exhibit similar differences, suggesting that this could be due to stochasticity in sampling these modes. The two standard NS runs also exhibit differences to each other in many of the other parameters too, another indication that any differences in the posteriors likely arise from the increased stochasticity of sampling so many modes. A clustered run with \textsc{Polychord} was performed for this example to better quantify the multi-modality, which reported $58$ clusters at the end of the run. 

\bsp	
\label{lastpage}
\end{document}

%% file: nested_sampling.pdf_tex
\begingroup%
  \makeatletter%
  \providecommand\color[2][]{%
    \errmessage{(Inkscape) Color is used for the text in Inkscape, but the package 'color.sty' is not loaded}%
    \renewcommand\color[2][]{}%
  }%
  \providecommand\transparent[1]{%
    \errmessage{(Inkscape) Transparency is used (non-zero) for the text in Inkscape, but the package 'transparent.sty' is not loaded}%
    \renewcommand\transparent[1]{}%
  }%
  \providecommand\rotatebox[2]{#2}%
  \newcommand*\fsize{\dimexpr\f@size pt\relax}%
  \newcommand*\lineheight[1]{\fontsize{\fsize}{#1\fsize}\selectfont}%
  \ifx\svgwidth\undefined%
    \setlength{\unitlength}{595.27559055bp}%
    \ifx\svgscale\undefined%
      \relax%
    \else%
      \setlength{\unitlength}{\unitlength * \real{\svgscale}}%
    \fi%
  \else%
    \setlength{\unitlength}{\svgwidth}%
  \fi%
  \global\let\svgwidth\undefined%
  \global\let\svgscale\undefined%
  \makeatother%
  \begin{picture}(1,1.41428571)%
    \lineheight{1}%
    \setlength\tabcolsep{0pt}%
    \put(0,0){\includegraphics[width=\unitlength,page=1]{nested_sampling.pdf}}%
    \put(0.39552759,0.89731083){\color[rgb]{0,0,0}\makebox(0,0)[lt]{\lineheight{1.25}\smash{\begin{tabular}[t]{l}$0$\end{tabular}}}}%
    \put(0,0){\includegraphics[width=\unitlength,page=2]{nested_sampling.pdf}}%
    \put(0.05982268,1.21366478){\color[rgb]{0,0,0}\makebox(0,0)[lt]{\lineheight{1.25}\smash{\begin{tabular}[t]{l}$\mathcal{L}_1$\end{tabular}}}}%
    \put(0.16891574,0.9963987){\color[rgb]{0,0,0}\makebox(0,0)[lt]{\lineheight{1.25}\smash{\begin{tabular}[t]{l}$\mathcal{L}_2$\end{tabular}}}}%
    \put(0.1890745,1.06443431){\color[rgb]{0,0,0}\makebox(0,0)[lt]{\lineheight{1.25}\smash{\begin{tabular}[t]{l}$\mathcal{L}_4$\end{tabular}}}}%
    \put(0.21679274,1.15514856){\color[rgb]{0,0,0}\makebox(0,0)[lt]{\lineheight{1.25}\smash{\begin{tabular}[t]{l}$\mathcal{L}_3$\end{tabular}}}}%
    \put(0.36834985,1.07621961){\color[rgb]{0,0,0}\makebox(0,0)[lt]{\lineheight{1.25}\smash{\begin{tabular}[t]{l}$\mathcal{L}$\end{tabular}}}}%
    \put(0.64049276,0.89857072){\color[rgb]{0.02352941,0.00392157,0.00392157}\makebox(0,0)[lt]{\lineheight{1.25}\smash{\begin{tabular}[t]{l}$X_1$\\\end{tabular}}}}%
    \put(0.7659437,0.92358211){\color[rgb]{0,0,0}\makebox(0,0)[lt]{\lineheight{1.25}\smash{\begin{tabular}[t]{l}$X$\end{tabular}}}}%
    \put(0.72348615,0.89731083){\color[rgb]{0,0,0}\makebox(0,0)[lt]{\lineheight{1.25}\smash{\begin{tabular}[t]{l}$1$\end{tabular}}}}%
    \put(0.56687632,0.89731083){\color[rgb]{0,0,0}\makebox(0,0)[lt]{\lineheight{1.25}\smash{\begin{tabular}[t]{l}$X_2$\\\end{tabular}}}}%
    \put(0.19698534,0.89731083){\color[rgb]{0,0,0}\makebox(0,0)[lt]{\lineheight{1.25}\smash{\begin{tabular}[t]{l}$\theta_2$\end{tabular}}}}%
    \put(-0.0125128,1.0836903){\color[rgb]{0,0,0}\makebox(0,0)[lt]{\lineheight{1.25}\smash{\begin{tabular}[t]{l}$\theta_1$\end{tabular}}}}%
    \put(0.494,0.89731083){\color[rgb]{0,0,0}\makebox(0,0)[lt]{\lineheight{1.25}\smash{\begin{tabular}[t]{l}$X_3$\\\end{tabular}}}}%
    \put(0.435,0.89731083){\color[rgb]{0,0,0}\makebox(0,0)[lt]{\lineheight{1.25}\smash{\begin{tabular}[t]{l}$X_4$\\\end{tabular}}}}%
    \put(0,0){\includegraphics[width=\unitlength,page=3]{nested_sampling.pdf}}%
  \end{picture}%
\endgroup%